\documentclass[sn-mathphys]{sn-jnl}

\usepackage{lineno}

\jyear{2023}%

\theoremstyle{thmstyleone}%
\theoremstyle{thmstyletwo}%

\theoremstyle{thmstylethree}%

\begin{document}

\title[Diffusion Models for Generation of Microscopy Image Data]{Denoising Diffusion Probabilistic Models for Generation of Realistic Fully-Annotated Microscopy Image Datasets}


\author*[1]{\fnm{Dennis} \sur{Eschweiler}}\email{dennis.eschweiler@lfb.rwth-aachen.de}

\author[1]{\fnm{Rüveyda} \sur{Yilmaz}}\email{rueveyda.yilmaz@lfb.rwth-aachen.de}

\author[1]{\fnm{Matisse} \sur{Baumann}}\email{matisse.baumann@lfb.rwth-aachen.de}

\author[1]{\fnm{Ina} \sur{Laube}}\email{ina.laube@lfb.rwth-aachen.de}

\author[1]{\fnm{Rijo} \sur{Roy}}\email{rijo.roy@lfb.rwth-aachen.de}

\author[1]{\fnm{Abin} \sur{Jose}}\email{abin.jose@lfb.rwth-aachen.de}

\author[1]{\fnm{Daniel} \sur{Brückner}}\email{daniel.brueckner@lfb.rwth-aachen.de}

\author*[1]{\fnm{Johannes} \sur{Stegmaier}}\email{johannes.stegmaier@lfb.rwth-aachen.de}

\affil[1]{\orgdiv{Institute of Imaging and Computer Vision}, \orgname{RWTH Aachen University}, \orgaddress{\city{Aachen}, \country{Germany}}}


\abstract{Recent advances in computer vision have led to significant progress in the generation of realistic image data, with denoising diffusion probabilistic models proving to be a particularly effective method. In this study, we demonstrate that diffusion models can effectively generate fully-annotated microscopy image data sets through an unsupervised and intuitive approach, using rough sketches of desired structures as the starting point. The proposed pipeline helps to reduce the reliance on manual annotations when training deep learning-based segmentation approaches and enables the segmentation of diverse datasets without the need for human annotations. This approach holds great promise in streamlining the data generation process and enabling a more efficient and scalable training of segmentation models, as we show in the example of different practical experiments involving various organisms and cell types.}





\maketitle
Enabling automated segmentation of cells in fluorescence microscopy image data is a crucial step in supporting biomedical experts in conducting a large variety of experiments~\cite{meijering2016,meijering2020}.
This variety in experimental settings is mirrored to the image data appearances, posing a challenge for segmentation approaches trained with the generally scarce variety of annotated image data.
To overcome this challenge, costly and tedious human annotations have to be acquired, causing a bottleneck in realizing the full potential of learning-based approaches and restricting their application in practice.
Annotation efforts are reduced by automated data augmentation approaches~\cite{shorten2019,zhao2019,eschweiler2022} and tweaked segmentation pipelines~\cite{zhou2021, isensee2021nnu}, which help to ease the challenge, but still demand a small set of fully-annotated image data as a basis.
Alternatively, automated simulation approaches replicate desired characteristics of cellular structures in arbitrary amounts of image data~\cite{svoboda2016,boehland2019,eschweiler2021,baehr2021,wiesner2022,bruch2023} and ideally serve as a way to entirely replace human annotation.

\begin{figure}[t]
    \centering
    \includegraphics[width=0.99\textwidth]{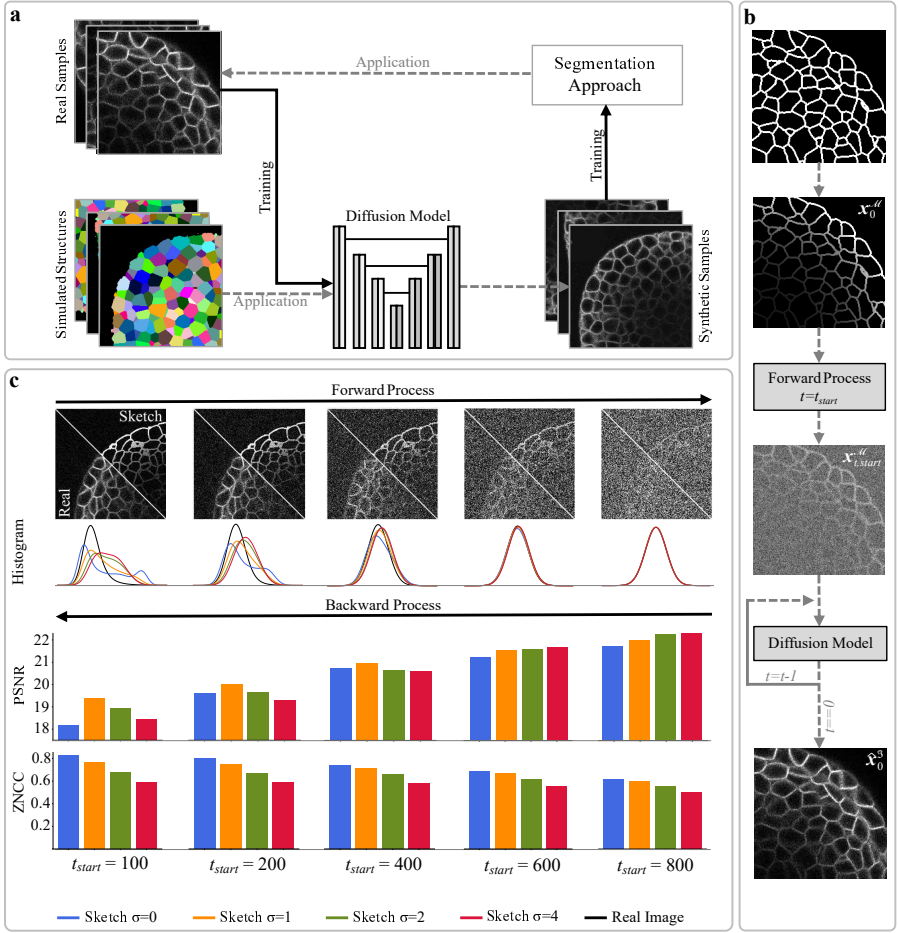}
    \caption{\footnotesize\textbf{Pipeline optimization.} (a) The whole pipeline involves training a diffusion model on real image data and applying it to simulated structures to generate fully-annotated image datasets, which are then used to train models that segment the real data. (b) During application of DDPMs, annotations are automatically turned into coarse sketches for a subsequent application of the forward process, to achieve a realistic generation of the corresponding image data. (c) Noisy data created by the forward process from either real images or sketches needs to be sufficiently similar to allow for the generation of realistic image data in the backward process. Peak signal-to-noise ratio (PSNR) and zero-normalized cross-correlation (ZNCC) are used as metrics, to assess realistic image data generated from different starting points $t_\mathrm{start}$ and sketch blurring factors $\sigma$.}
    \label{fig:optimizing}
\end{figure}
Recently, denoising diffusion probabilistic models (DDPM)~\cite{ho2020} have shown great potential in generating realistic image data, while neither requiring annotated training data, nor adversarial training concepts, as opposed to commonly used generative adversarial networks (GAN)~\cite{boehland2019,eschweiler2021,bruch2023}.
We propose a pipeline, which involves DDPM for the automated generation of fully-annotated image data, which is used to subsequently train segmentation approaches (Fig.\ref{fig:optimizing}a).
In DDPM, a gradual forward process is defined as a Markov chain, iteratively adding a small amount of noise to a real image $\mathbf{x}_0$ until reaching pure noise $\mathbf{x}_\mathrm{T}$~\cite{ho2020}.
A corresponding backward process is defined in which a neural network is trained to iteratively reverse the forward process, leading to the generation of realistic image data $\mathbf{\Hat{x}}_0$ from $\mathbf{x}_\mathrm{T}$~\cite{nichol2021}.
However, starting from pure noise $\mathbf{x}_\mathrm{T}$ does not allow to generate fully-annotated datasets due to the lack of control over the generated structures and the absence of corresponding annotations.
To address those issues, two adaptions are made to the application of DDPM.
First, the backward process is initiated early at $t_\mathrm{start}<\mathrm{T}$, ensuring that a significant portion of structural indications in $\mathbf{x}_{\mathrm{t,start}}$ is not yet fully obscured by noise and can be preserved throughout the generation of realistic image data~\cite{meng2021}.
Second, to allow for an intuitive modelling and control over cellular structures within the generated image data, sketches replace real image data as a starting point for the forward process generating $\mathbf{x}_{\mathrm{t,start}}$. 
These sketches can be simulated and provide indications of cell positions, shapes, and coarse structural characteristics, specifying a brief outline of the desired scene to be generated.
As the learned backward process was solely trained on real image data, the subsequent application to $\mathbf{x}_{\mathrm{t,start}}$ results in the generation of corresponding realistic image data (Fig.\ref{fig:optimizing}b).
Ultimately, with the known cell outline and positioning within the sketches, the pipeline is able to generate realistic fully-annotated image data in a automated and unsupervised manner.
However, in order to maintain high realism during the backward process, it is crucial for the noisy samples $\mathbf{x}_{\mathrm{t,start}}$ originating from the sketch domain $\mathcal{M}$ to exhibit data distributions that closely resemble those originating from the real image domain $\mathcal{I}$.
This poses an optimization challenge in determining the optimal value for the parameter $t_\mathrm{start}$.

Optimizing $t_\mathrm{start}$ requires balancing the generation of fine-grained details with the preservation of structural correlation to sketches.
To generate fine details, a substantial noise content from later stages of the forward process is necessary, but this can compromise the preservation of structural indications, which requires stopping the forward process as early as possible.
The details of this trade-off were evaluated using a publicly available fully-annotated 3D microscopy image dataset~\cite{willis2016}, which provides manually corrected annotations that enabled a precise assessment of various aspects of the proposed pipeline.
To analyze the forward process, both $\mathbf{x}_{\mathrm{t,start}}^\mathcal{M}$ and $\mathbf{x}_{\mathrm{t,start}}^\mathcal{I}$ were generated, and their similarity was measured by constructing data distribution histograms.
Additionally, to assess the learned backward process, all noisy samples were used to generate realistic $\mathbf{\Hat{x}}_0$, which were quantitatively evaluated using the peak signal-to-noise ratio (PSNR) for textural authenticity and the zero-normalized cross-correlation (ZNCC) for structural preservation (Fig.\ref{fig:optimizing}c).
Furthermore, we found that applying a Gaussian smoothing with standard deviation $\sigma$ to sketches before applying the forward process helped to prevent unnaturally sharp edges and reach similar data distributions earlier, allowing to choose an earlier $t_\mathrm{start}$.
In general, the optimal value of $t_\mathrm{start}$ is preferred to be set as early as possible to ensure the maximum structural correlation between sketches and image data. 
Additionally, since the generation process is iterative, selecting an earlier $t_\mathrm{start}$ directly translates to lower generation times.
We empirically determined $t_\mathrm{start}=400$ and $\sigma=1$ to offer a good trade-off between generative capacity and structural preservation.
\begin{figure}[h!]
    \centering
    \includegraphics[width=0.99\textwidth]{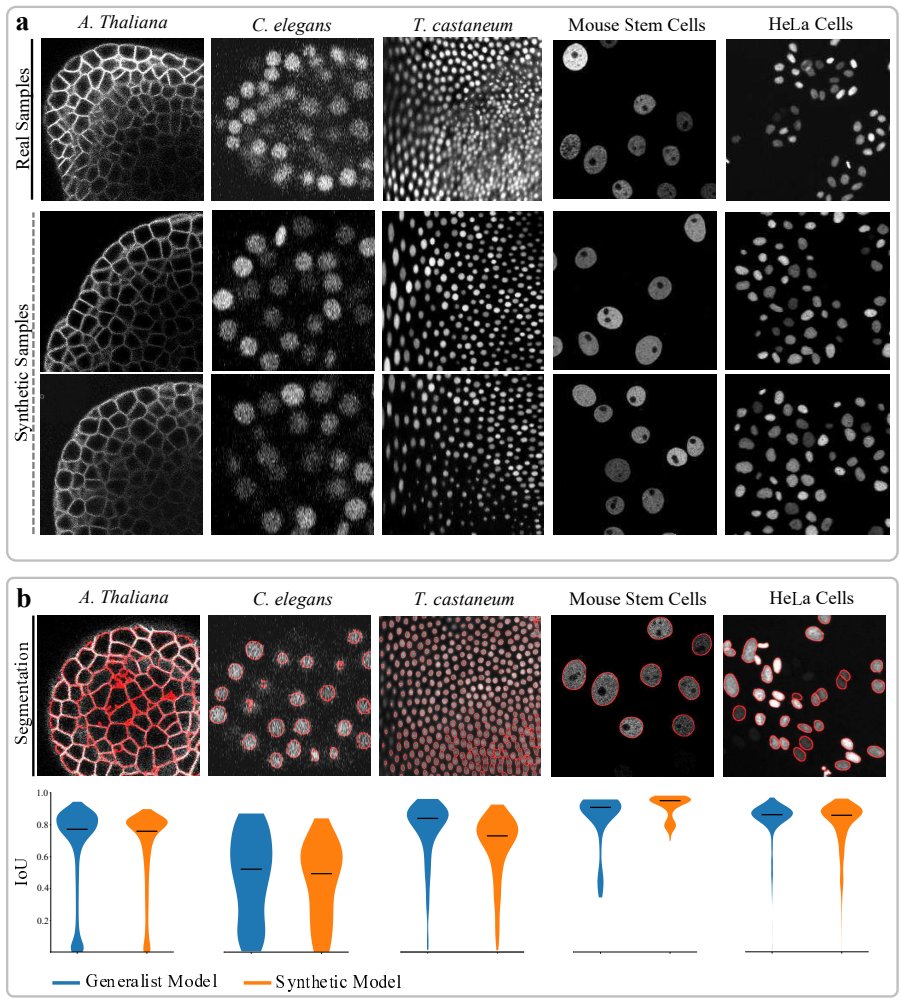}
    \caption{\footnotesize\textbf{Application} (a) Real image samples and fully-synthetic image samples generated by the diffusion model using simulated structures. (b) The Cellpose segmentation approach~\cite{stringer2020} is trained on synthetic datasets and applied to real image data to generate results (red overlay) without requiring human-generated annotations. Intersection-over-Union (IoU) scores obtained for a publicly available generalist model trained on a large collection of manually annotated image data (blue) and the model trained on synthetic data (orange) are shown as violin plots with indications of median values (black bar). All datasets are publicly available from~\cite{willis2016,ulman2017,faure2016,moreno2022,eschweiler2021}.}
    \label{fig:data}
\end{figure}

While manually annotated structures were used during the optimization of the parameters to test the generative aspects of the pipeline without the influence of structural differences, using manually annotated data does not represent a practical scenario for generating fully-annotated image datasets.
To assess different strategies for automation in the pipeline, rough segmentations were employed as sources to obtain sketches for various organisms and cell types.
Publicly available generalist segmentation approaches~\cite{stegmaier14a,eschweiler2021b} and publicly available silver truth annotations~\cite{ulman2017} were used to obtain rough representations of cell shapes for nuclei in \textit{C.~elegans}~\cite{murray2008,ulman2017}, \textit{D.~rerio}~\cite{faure2016}, mouse stem cells~\cite{bartova2011,ulman2017}, HeLa cells~\cite{neumann2010,ulman2017,moreno2022} and for cellular membranes in \textit{D.~rerio}~\cite{faure2016}.
Image quality was assessed in regions where annotations were available using the PSNR as a metric, with mean scores ranging between 19.58\,dB and 29.97\,dB across all data sets.
Although the annotations may contain errors that could affect the reported quality scores, they do not have a significant impact on the application of the proposed pipeline.
This is because the generated image data is directly correlated to the structures present in the annotations, enabling the creation of error-free image-mask pairs.

Relying on segmentations for collecting realistic structures can limit the scalability of the pipeline, as it is constrained by the availability of accurate generalist segmentation approaches.
On the other hand, simulations present a more complex scenario but offer greater potential for generalizability and scalability.
The main challenge lies in finding a simulation technique that can accurately reproduce the structural features visible in real image data, as any inaccuracies could introduce a domain gap between real and final synthetic image data.
Despite these challenges, conducting experiments with simulation approaches allows for exploring the full potential and limits of the pipeline.
As directly assessing the realism of the image data generated from simulated sketches was challenging due to the absence of corresponding real image data, we followed a more practical way of evaluation.
Instead, we focused on determining the usability of the generated data as training data by training the Cellpose approach~\cite{stringer2020} from scratch, followed by its application to real image data.
By using the accuracy of the segmentation results as a proxy, the realism of the generated data is indirectly assessed.
Additionally, segmentation results were compared to those obtained by the publicly available pretrained Cellpose model, which served as a baseline and a reference for models trained on a large, diverse and fully-annotated image dataset.
Simulations were obtained for five different datasets including cellular membranes in \textit{A.~thaliana}, and nuclei in \textit{C.~elegans}~\cite{murray2008,ulman2017}, \textit{T.~castaneum}~\cite{ulman2017}, mouse stem cells~\cite{bartova2011,ulman2017} and HeLa cells~\cite{neumann2010,ulman2017}~(Fig.\ref{fig:data}a).
The results presented in Figure~\ref{fig:data}b demonstrate that both models perform comparably well for each dataset, despite the fact that the models trained solely on synthetic image data were trained with only 200 generated samples.
It is noteworthy that the synthetic model achieved comparable scores without the need for any human-generated annotations, surpassing the requirement for a large collection of annotated image data as in the generalist model~\cite{stringer2020}.
This capability enables the potential application of segmentation models to entirely different datasets and structures, where generalist segmentation models would typically encounter challenges or limitations.
With sparse ground truth data available for evaluation, a Wilcoxon rank-sum test~\cite{mann1947} was conducted to identify potential differences in segmentation accuracy between the models.
With the largest $p$-value reaching 0.0002 for the dataset showing mouse stem cells, the test confirmed that the performance of both models on all datasets is comparable.
This suggests the additional potential to use the generated data for systematic evaluation and benchmarking purposes to enable the evaluation of segmentation models on specific datasets without requiring any manual annotations.

Overall, the consistently high PSNR values for the synthetic image data and the segmentation results comparable to state-of-the-art approaches trained with large annotated datasets emphasize the realism of the generated data and demonstrate its practical usability.
However, there is one limitation to this approach, as the noise introduced during the forward process makes it challenging to produce very dark cells, as the structural information gets lost in the added noise.
Consequently, dimly illuminated regions pose difficulties for accurate segmentation as the models trained on synthetic data may not be fully-equipped to handle all challenges posed by real data.
Despite this limitation, the obtained segmentation scores demonstrate the capability of training specialized models that achieve state-of-the-art segmentation results in a fully-automated manner with data generated by an intuitive and unsupervised approach.
The pipeline addresses the challenge of manual annotation requirements for applying segmentation approaches and suggests shifting the focus towards identifying suitable setups for creating cellular structures.
The generation of these structures can be achieved through simulation approaches or by utilizing publicly available generalist segmentation methods like Cellpose~\cite{stringer2020} or SAM~\cite{kirillov2023}.
Consequently, the application of deep learning-based segmentation approaches becomes more accessible for datasets with limited and absent annotations.
To contribute towards this goal, all fully-annotated synthetic image datasets are publicly available at \url{https://osf.io/dnp65/}, and code for training and application is available at \url{https://github.com/stegmaierj/DiffusionModelsForImageSynthesis}.

\clearpage
\backmatter
\newpage


\bibliography{main}


\end{document}